\documentclass[]{imsart}
\usepackage{amsthm,amsmath}
\usepackage[colorlinks,citecolor=blue,urlcolor=blue]{hyperref}

\usepackage{amsmath,amsfonts,amssymb,amsthm}

\usepackage{enumitem}

\RequirePackage{amsthm,amsmath,amsfonts,amssymb}
\RequirePackage[numbers]{natbib}
\RequirePackage[colorlinks,citecolor=blue,urlcolor=blue]{hyperref}
\RequirePackage{graphicx}
\usepackage{lineno,hyperref}

\startlocaldefs
\modulolinenumbers[5]

\usepackage[utf8]{inputenc}
\usepackage[english]{babel}

\usepackage[margin=1in]{geometry}
\usepackage{bm}
\usepackage{graphicx}
\usepackage{amssymb,amsmath,amsthm,mathrsfs}
\usepackage{epstopdf}
\usepackage{algorithm}
\usepackage{amsfonts,delarray}
\usepackage{algorithm,algcompatible,amsmath}
\usepackage{rotating,color}
\usepackage{subfigure}
\usepackage{multirow}
\usepackage{titlesec,textcase}
\usepackage{longtable}
\usepackage{bbm}
\usepackage{rotating}

\newtheorem{Theorem}{Theorem}[section]
\newtheorem{Lemma}[Theorem]{Lemma}
\newtheorem{Corollary}[Theorem]{Corollary}

\theoremstyle{definition}

\RequirePackage[normalem]{ulem} 

\DeclareMathOperator*{\argmin}{arg\,min}
\DeclareMathOperator*{\argmax}{arg\,max}

\definecolor{rp}{RGB}{83,54,106}

\def\boxit#1{\vbox{\hrule\hbox{\vrule\kern6pt\vbox{\kern6pt#1\kern6pt}\kern6pt\vrule}\hrule}}

\endlocaldefs

\begin{document}
\begin{frontmatter}
\title{Information Limits for Detecting a Subhypergraph}

\runtitle{Detection of a Subhypergraph}
\runauthor{ M. Yuan and Z. Shang}
\begin{aug}
\author[A]{\fnms{Mingao} \snm{Yuan}\ead[label=e1]{mingao.yuan@ndsu.edu}}
\and
\author[B]{\fnms{Zuofeng} \snm{Shang}\ead[label=e2]{zshang@njit.edu}}
\address[A]{Department of Statistics,
North Dakota State University,
\printead{e1}}

\address[B]{Department of Mathematical Sciences,
New Jersey Institute of Technology,
\printead{e2}}
\end{aug}

\begin{abstract}
We consider the problem of recovering a subhypergraph based on an observed adjacency tensor corresponding to a uniform hypergraph. 
The uniform hypergraph is assumed to contain a subset of vertices called as subhypergraph. The edges restricted to the subhypergraph are assumed to follow a different probability distribution than other edges. We consider both weak recovery and exact recovery of the subhypergraph, and establish information-theoretic limits in each case. Specifically, we establish sharp conditions for the possibility of weakly or exactly recovering the subhypergraph from an information-theoretic point of view.
These conditions are fundamentally different from their counterparts derived in hypothesis testing literature.
\end{abstract}



\end{frontmatter}

\section{Introduction}
\label{S:1}
An \textit{undirected} $m$-uniform hypergraph is a pair $(\mathcal{V},\mathcal{E})$ in which $\mathcal{V}=[N]:=\{1,2,\dots,N\}$ 
is a vertex set and $\mathcal{E}$ is an edge set. Each edge in $\mathcal{E}$ is consists of exactly $m$ vertices in $\mathcal{V}$. 
The corresponding adjacency tensor is an $m$-dimensional symmetric array $A\in(B^N)^{\otimes m}$ 
satisfying $A_{i_1i_2\ldots i_m}\in B$ for $1\leq i_1<i_2<\dots<i_m\leq N$, in which $B\subset\mathbb{R}$. 
Here, symmetry means that $A_{i_1i_2\ldots i_m}=A_{j_1j_2\ldots j_m}$ whenever $i_1,i_2,\ldots,i_m$ is a permutation of $j_1,j_2,\ldots,j_m$. 
If $|\{i_1,i_2,\ldots,i_m\}|\leq m-1$, then $A_{i_1i_2\ldots i_m}=0$,
i.e., no self-loops are allowed. 
In particular, $B=\{0,1\}$ corresponds to binary hypergraphs.
The general $B$ corresponds to weighted hypergraphs.

Given probability distributions $P$ and $Q$ over $B$, let $\mathcal{H}_m(N,Q; n,P)$ denote a uniform hypergraph model including a
subhypergraph of cardinality $n$ defined as follows: for a uniformly and randomly drawn subset $S^*\subset [N]$ with $|S^*|=n$, 
\begin{equation}\label{our:model}
A_{i_1i_2\ldots i_m}\sim \left\{
\begin{array}{cc}
    P,& i_1,\dots, i_m\in S^*,  \\
    Q, & \textrm{otherwise.}
\end{array}
\right.
\end{equation}
Observing $A$, we are interested in recovering
the subhypergraph $S^*$.
When $P$ and $Q$ are both Bernoulli distributions, various algorithms have been developed by \cite{AC09,C00, CS10, GKT05, HWX18,KS09,CKKMP18,WLKN09} for  $m=2$, and by \cite{HWC17,T15,LJY15,ZHS06,CDKKR18,BGK17,LZ20,K11,LJY15} for $m\geq3$. For general $P$ and $Q$ with $m=2$, \cite{HWX17} studied this problem systematically from information-theoretic viewpoint and derived sharp recovery boundaries. For $m\geq3$, relevant literature only exist 
in block models including \cite{ALS18, ALS19, CLW18, LKH21}. Specifically,
under stochastic block models, \cite{CLW18} obtained minimax detection bounds,
\cite{ALS18} derived sufficient conditions for weak or exact recovery,
\cite{LKH21} derived exact recovery information limits;
 under generalized censored block models, \cite{ALS19} derived sufficient and necessary conditions for weak recovery.
To the best of ourknowledge,
information limits for detecting a subhypergraph under model (\ref{our:model}) are nonexistent. In this paper, we will study this problem for arbitrary probability distributions $P,Q$ and integer $m\geq2$. 

Our theoretical results include
information limits for weak/exact recovery of
the subhypergraph obtained in an asymptotic regime
$N\rightarrow\infty$ and $\lim\sup n/N<1$. 
Let $D(P\|Q)=\mathbb{E}_P\left(\log\frac{dP}{dQ}\right)$ be the Kullback-Leibler divergence from $Q$ to $P$ and both $D(P\|Q)$, $D(Q\|P)$ are supposed to be finite. 
We derive sharp regions characterized by $N,n$ for
weakly or exactly recovering the subhypergraph with computational complexity aside which provides a benchmark for developing polynomial time algorithms. Specifically, our main results are summarized in Table \ref{region}.
\begin{table}[H]
\vspace{-2mm}
	\centering
		\caption{\it Regions for recovering a subhypergraph.}
	\begin{tabular}{|p{8cm} p{6cm} |}
	\hline
	Region & Detectibility   \\
	\hline
	(a) $n^{m-1}D(P\|Q)\rightarrow\infty$ and $\liminf_{N\rightarrow\infty}\frac{n^{(m)}D(P\|Q)}{n\log\frac{N}{n}}>1$ & Weak recovery of subhypergraph is possible        \\
 (b) $n^{m-1}D(P\|Q)=O(1)$ or $\limsup_{N\rightarrow\infty}\frac{n^{(m)}D(P\|Q)}{n\log\frac{N}{n}}<1$   &	 Weak recovery of subhypergraph is impossible           \\
 \hline
 	(a) and $\liminf_{N\rightarrow\infty}\frac{n^{(m-1)}E_Q\left(\frac{1}{n^{(m-1)}}\log\frac{N}{n}\right)}{\log N}>1$   &	 Exact recovery of subhypergraph is possible        \\
	(b) or $\limsup_{N\rightarrow\infty}\frac{n^{(m-1)}E_Q\left(\frac{1}{n^{(m-1)}}\log\frac{N}{n}\right)}{\log N}<1$    & Exact recovery of subhypergraph is impossible          \\
	\hline
\end{tabular}
\label{region}
\vspace{-5mm}
\end{table}

\begin{figure}[htp] 
\vspace{-3mm}
\centering
\includegraphics[scale=0.4]{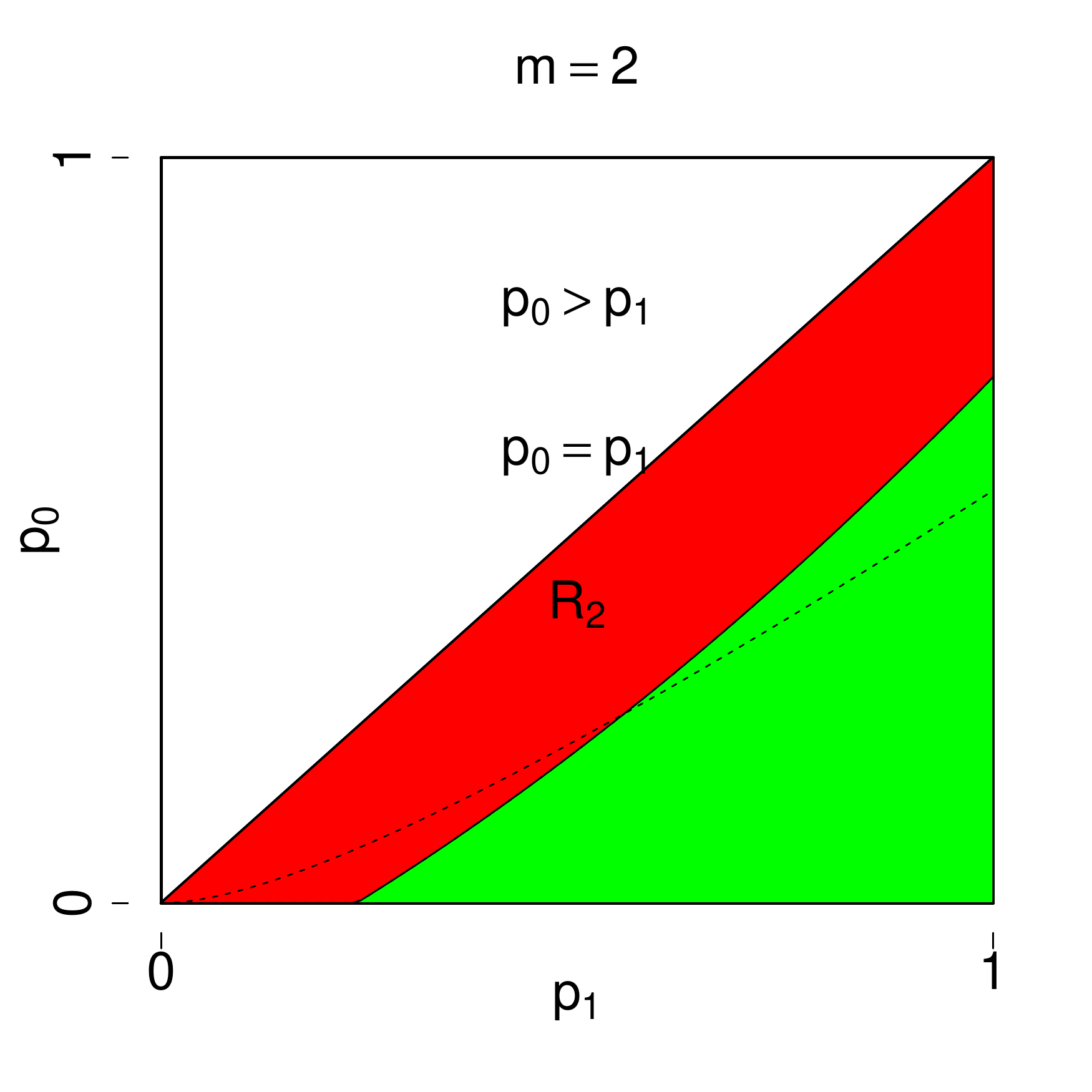}
\hspace{20mm}
\includegraphics[scale=0.4]{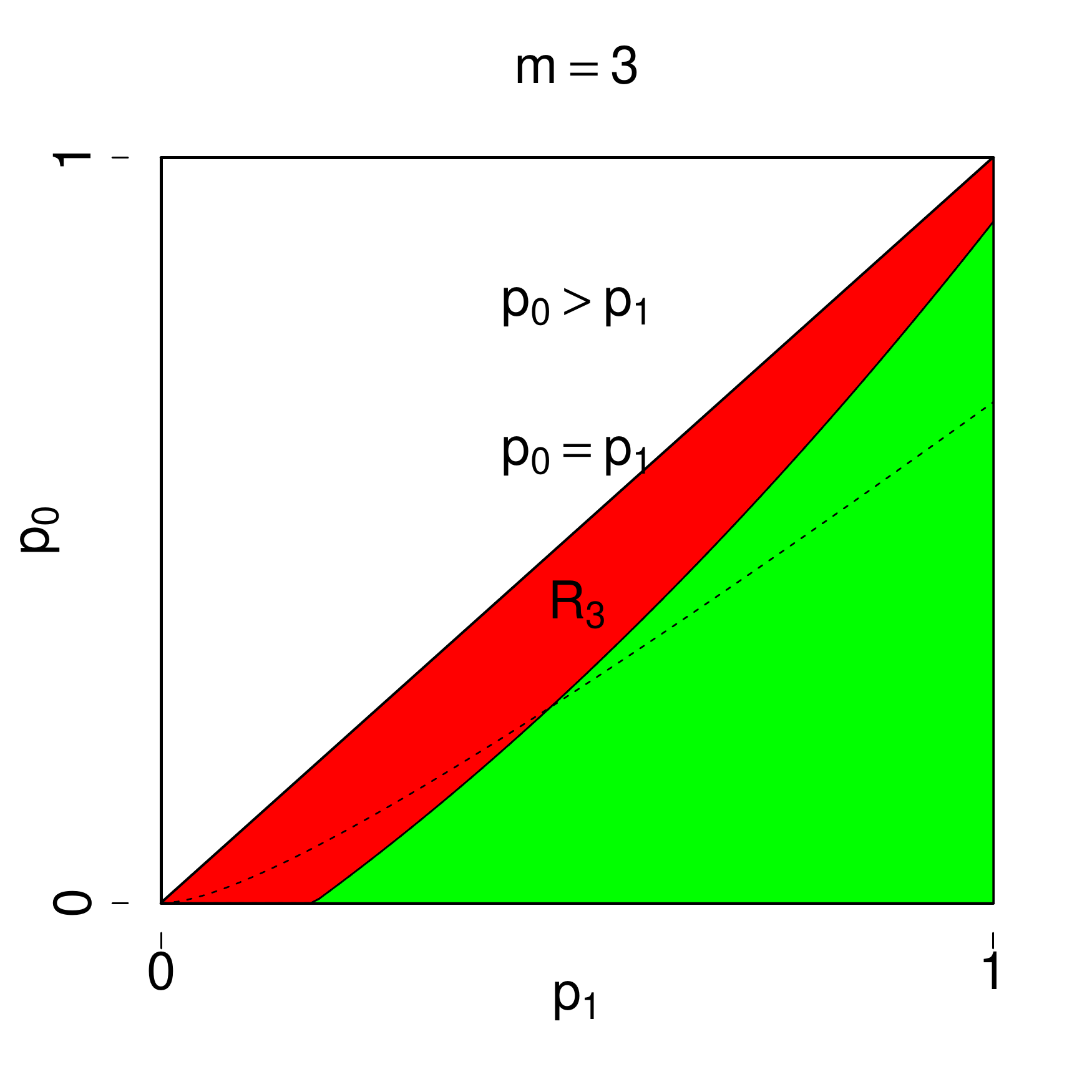}
\vspace{-5mm}
\caption{\it\small Week recovery regions in terms of $(p_1,p_0)$ for  $m=2,3$. Red: week recovery is impossible; green: weak recovery is possible. }
\label{algsubHperphase}
\end{figure}

\begin{figure}[h] 
\vspace{-3mm}
\centering
\includegraphics[scale=0.4]{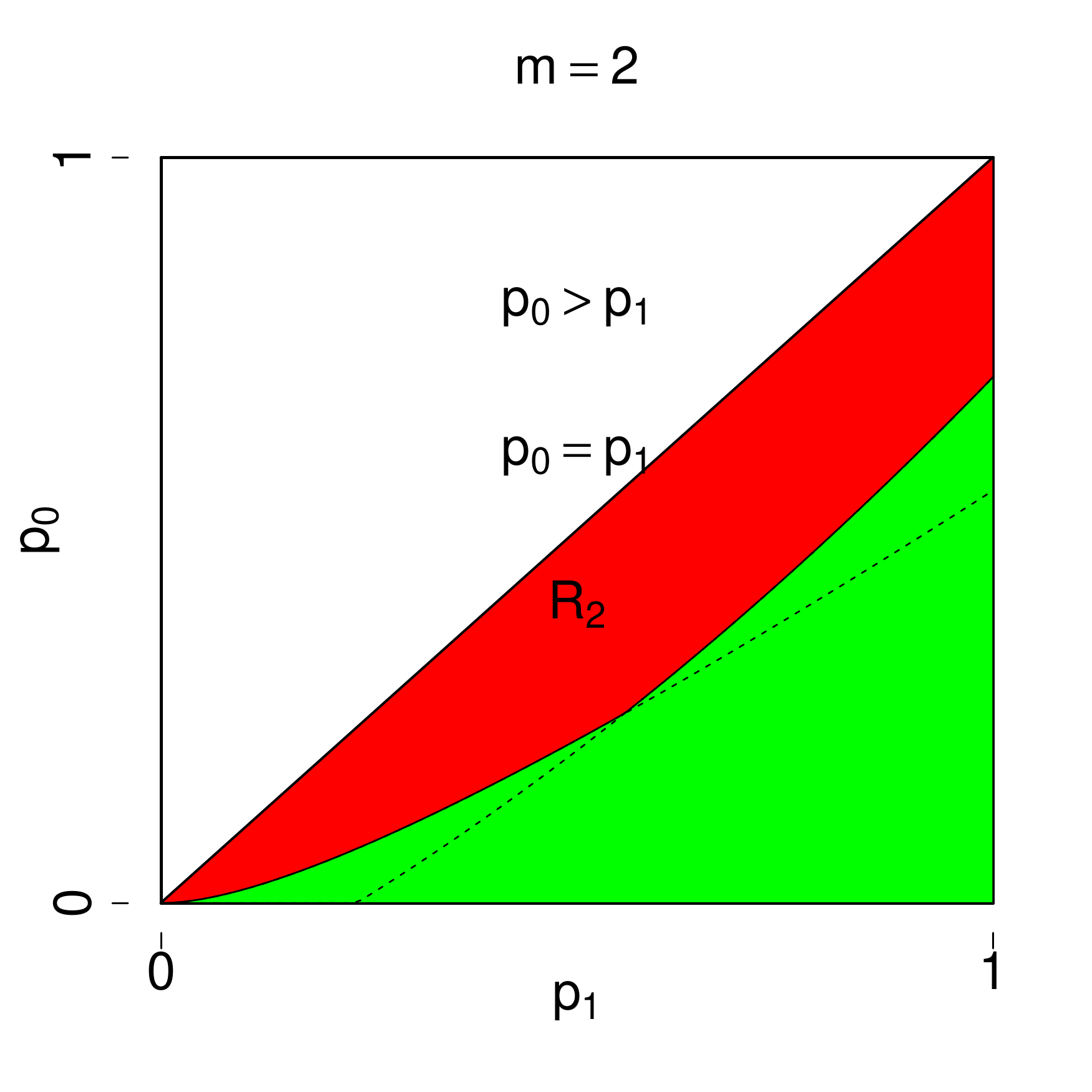}
\hspace{20mm}
\includegraphics[scale=0.4]{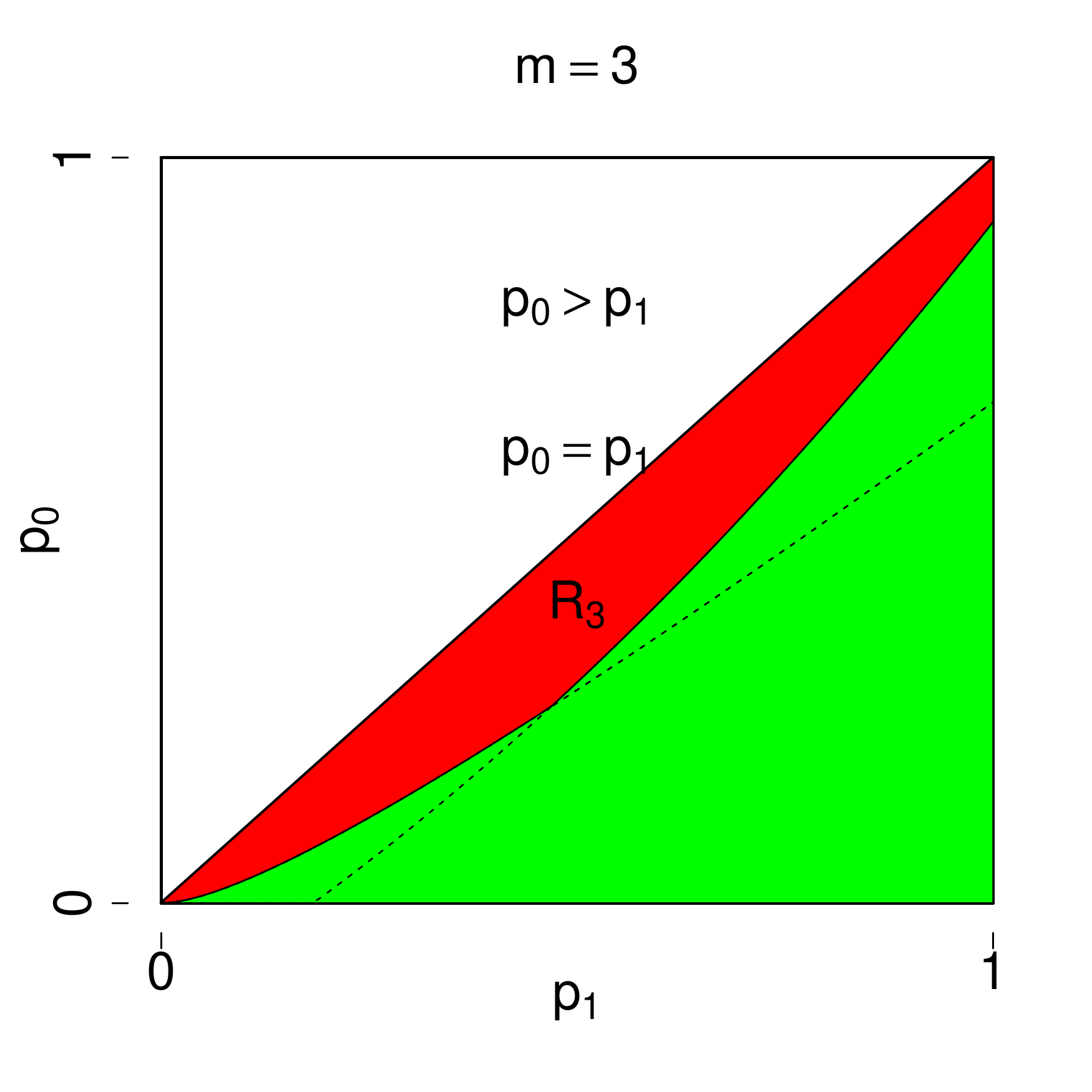}
\vspace{-5mm}
\caption{\it\small Hypothesis testing regions in terms of $(p_1,p_0)$ for $m=2,3$. Red: all tests are powerless; green: a powerful test exists. }
\label{subHperphase}
\end{figure}

In the special case where $P=\textrm{Bern}(p_1)$ and $Q=\textrm{Bern}(p_0)$ are both Bernoulli distributions, we 
are able to visualize the weak recovery region characterized by $(p_1,p_0)$ for $m=2,3$ in Figure \ref{algsubHperphase}. Red (green) region indicates that weak recovery is impossible (possible). Clearly, the red region corresponds to $m=3$ is smaller than the one corresponding to $m=2$, indicating that detecting a subhypergraph is easier than detecting a graph. \cite{YS21,YS21b} make a similar discovery in hypothesis testing scenario, i.e.,
testing the existence of subhypergraph is easier than testing the existence of a subgraph. Figure \ref{subHperphase} displays the regions characterized by $(p_1,p_0)$ for testing the existence of a subhypergraph for $m=2,3$;
red (green) region indicates that a powerful test is impossible (possible).
Interestingly, for fixed integer $m$, the red region in Figure \ref{subHperphase} is smaller than the red region in Figure \ref{algsubHperphase}, indicating that hypothesis testing problem is generally easier to solve than detection problem.

\section{Main Results}\label{subgraph}
Let $Z=(Z_1,\ldots,Z_N)$ denote the membership vector of $S^*$, i.e.,
$Z_i=1$ if $i\in S^*$ and $Z_i=0$ otherwise. 
For $\widehat{S}$ an estimator of $S^*$,
let $\widehat{Z}$ denote the membership vector of $\widehat{S}$. We say that $\widehat{Z}$ is an \textit{exact  recovery} of $Z$ if $\mathbb{P}(\widehat{Z}\neq Z)=o(1)$; $\widehat{Z}$ is a \textit{weak recovery} of $Z$ if $\mathbb{P}(d_H(\widehat{Z},Z)/n)=o(1)$,
where $d_H(\widehat{Z},Z)$ is the Hamming distance between $\widehat{Z}$
and $Z$. Obviously, when $n=O(1)$, weak recovery and exact recovery are equivalent. 
Let $L=\log\frac{dP}{dQ}$ and $\phi_Q(x)=\log\mathbb{E}_Q[e^{xL}]$. Moreover, assume that $P$ and $Q$ satisfy the following property:
\begin{equation}\label{cond:1} 
\frac{d^2}{dx^2}\phi_Q(x)=O\left(\min\left\{D(P\|Q),D(Q\|P)\right\}\right),\ \ x\in [-1,1].
\end{equation}
This condition requires the log moment generating functions $\phi_Q(x)$ of $L$ is bounded by the divergences on the interval $[-1,1]$. 
The following theorem provides a sufficient condition under which
weak recovery is possible.
\begin{Theorem}[Weak Recovery]\label{thm:1}
Suppose that $P$ and $Q$ satisfy (\ref{cond:1}). If
\begin{equation}\label{cond:2} 
n^{m-1}D(P\|Q)\rightarrow\infty,\ \ \ \ \ \ \ \ \liminf_{N\rightarrow\infty}\frac{n^{(m)}D(P\|Q)}{n\log\frac{N}{n}}>1,
\end{equation}
then there exists an estimator $\widehat{S}$ of $S^*$ such that
\begin{equation}\label{weak} 
\mathbb{P}\left[|\widehat{S}\Delta S^*|\leq 2n\epsilon_m\right]\geq 1-e^{-\Omega\left(\frac{n}{\epsilon_m}\right)},\ \ \textrm{where}\,\, \epsilon_m=\left(n^{(m-1)}D(P\|Q)\right)^{-\frac{1}{2}}.
\end{equation}
Conversely, if there is an estimator $\widehat{S}$ such that $\mathbb{E}[d_H(\widehat{Z},Z)]=o(n)$, then \begin{equation}\label{cond:3} 
n^{m-1}D(P\|Q)\rightarrow\infty,\ \ \ \ \ \ \ \ \liminf_{N\rightarrow\infty}\frac{n^{(m)}D(P\|Q)}{n\log\frac{N}{n}}\geq1.
\end{equation}
\end{Theorem}
 When $D(P\|Q)$ remains constant, condition (\ref{cond:2}) is weaker for $m=3$ than $m=2$. The later has been considered for weak recovery of subgraph by \cite{HWX17}. Besides, the error rate $\epsilon_m$ in (\ref{weak}) is smaller for $m=3$ than $m=2$. This indicates the significant difference between subhypergraph recovery and subgraph recovery.
 
 Theorem \ref{thm:1} says that (\ref{cond:2}) are sufficient conditions
 for weak recovery. Meanwhile, (\ref{cond:2}) are almost necessary conditions for weak recovery, since if (\ref{cond:3}) fails,
 then weak recovery is impossible.
 These conditions are derived under model (\ref{our:model}),
 and are dramatically different from the ones derived under 
 stochastic block models (\cite{CLW18,ALS18,LKH21})
 or generalized censored block models (\cite{ALS19}).
 The theorem holds for either $n=o(N)$ or $n\asymp N$, i.e.,
 the cardinality of the underlying subhypergraph is either significantly smaller than $N$ or of the same order as $N$.

Theorem \ref{thm:1} is established in the perspective of weak recovery, fundamentally different from the ones derived in hypothesis testing (\cite{YS21}). To demonstrate the difference,
consider unweighted hypergraphs with $P=\textrm{Bern}(p_1)$ and $Q=\textrm{Bern}(p_0)$. Define $H_{p}(q)=q\log\frac{q}{p}+(1-q)\log\frac{1-q}{1-p}$ for $p, q\in (0,1)$. Then it can be seen that $D(P\|Q)=H_{p_0}(p_1)$.
By Theorem \ref{thm:1}, we have the following corollary.
\begin{Corollary}\label{weakcor}
Suppose $P=\textrm{Bern}(p_1)$ and $Q=\textrm{Bern}(p_0)$, $\log\frac{p_1}{p_0}$ and $\log\frac{1-p_1}{1-p_0}$ are bounded. Then weak recovery is possible if
\begin{equation}\label{corcond:1} 
n^{m-1}H_{p_0}(p_1)\rightarrow\infty,\ \ \ \ \ \ \ \ \liminf_{N\rightarrow\infty}\frac{n^{(m)}H_{p_0}(p_1)}{n\log\frac{N}{n}}>1.
\end{equation}
Conversely, if weak recovery is possible, then
 \begin{equation}\label{corcond:2} 
n^{m-1}H_{p_0}(p_1)\rightarrow\infty,\ \ \ \ \ \ \ \ \liminf_{N\rightarrow\infty}\frac{n^{(m)}H_{p_0}(p_1)}{n\log\frac{N}{n}}\geq1.
\end{equation}
\end{Corollary}

It might be interesting to compare the weak recovery condition (\ref{corcond:1}) with the detection boundary condition in Theorem 3.1 of \cite{YS21}. 
Assume $n=o(N)$ as in \cite{YS21}, then by Corollary \ref{weakcor}, weak recovery is impossible if
\begin{equation}\label{comp}
    \liminf_{N\rightarrow\infty}\frac{n^{(m)}H_{p_0}(p_1)}{n\log\frac{N}{n}}<1.
\end{equation}
Note that (\ref{comp}) is weaker than Condition (7) in \cite{YS21}.
Hence, the undetectable region derived in \cite{YS21} is smaller than the weak recovery region for any integer $m\geq2$. We visualize the regions in Figure \ref{algsubHperphase} and Figure \ref{subHperphase}, which demonstrates the significant distinction between weak recovery limits and hypothesis testing.

 The proof of sufficiency in Theorem \ref{thm:1} proceeds by showing the maximum likelihood ratio estimator is a weak recovery estimator.
Let 
\[
L_{i_1\ldots i_m}=\left\{\begin{array}{cc}\frac{dP}{dQ}(A_{i_1\ldots i_m}),
&\textrm{if $i_1,\ldots, i_m\in\mathcal{V}$ are pairwise distinct},\\
0,&\textrm{otherwise.}
\end{array}\right.
\]
For any two subsets $S_1,S_2\subset\mathcal{V}$, let 
\[
I_m(S_1,S_2)=\{(i_1,\dots,i_m)| i_1<\dots<i_m,\{i_1,\dots,i_m\}\cap S_1\neq \emptyset,\{i_1,\dots,i_m\}\cap S_2\neq \emptyset, \{i_1,\dots,i_m\}\subset S_1\cup S_2\}
\]
and
\[
L(S_1,S_2)=\sum_{(i_1,\dots,i_m)\in I_m(S_1,S_2)}L_{i_1\dots i_m}.
\]
Clearly, $I_m(S_1,S_2)$ includes the tuples $(i_1,\ldots,i_m)$
that intersects with both both $S_1$ and $S_2$ 
whose entries are included in $S_1\cup S_2$.
Let $\widehat{S}_{ML}$ be the maximum likelihood estimator of $S^*$ defined as
\begin{equation}\label{MLE} 
\widehat{S}_{ML}=\argmax_{S\subset [N],|S|=n}L(S,S).
\end{equation}

\begin{proof}[Proof of Theorem \ref{thm:1}] 
{\bf (Sufficiency).} Let $\widehat{S}=\widehat{S}_{ML}$ and $R=|\widehat{S}\cap S^*|$. Then $|\widehat{S}|=|S^*|=n$ and $|\widehat{S}\Delta S^*|=2(n-R)$. To prove (\ref{weak}), we only need to show $\mathbb{P}\left(R\leq (1-\epsilon_m) n\right)\leq \exp\left(-\Omega\left(\frac{n}{\epsilon_m}\right)\right)$.

By condition (\ref{cond:2}), there exists a constant $\eta\in (0,1)$ such that $K(m,n)D(P\|Q)\geq (1-\eta)\log\frac{N}{n}$. Let $\theta=(1-\eta)D(P\|Q)$. For $0\leq r\leq n-1$, we have
\begin{eqnarray*}
\{R=r\}&\subset&\{\exists S\subset S^*: |S|=n-r, L(S,S^*)\leq (n^{(m)}-r^{(m)})\theta\}\\
&\cup& \{\exists S\subset S^*, \exists T\subset (S^*)^c: |S|=|T|=n-r, L(T,T)+L(T, S^*\setminus S)\geq (n^{(m)}-r^{(m)})\theta\}.
\end{eqnarray*}
Note that $A_{i_1\dots i_m}\sim P$ and  $A_{i_1\dots i_m}\sim Q$ for $\{i_1,\dots, i_m\}\subset S\subset S^*$ and $\{i_1,\dots, i_m\}\cap T\neq \emptyset$ respectively. Hence, we get
\begin{eqnarray*}
&&\mathbb{P}[R=r]\\
&\leq& \binom{n}{n-r}\mathbb{P}\left(L(S,S^*)\leq (n^{(m)}-r^{(m)})\theta\right)\\
&&+\binom{n}{n-r}\binom{N-n}{n-r}\mathbb{P}\left(L(T,T)+L(T, S^*\setminus S)\geq (n^{(m)}-r^{(m)})\theta\right)\\
&\leq&\left(\frac{ne}{n-r}\right)^{n-r}\exp\left(-(n^{(m)}-r^{(m)})E_P(\theta)\right)+\left(\frac{(N-n)ne^2}{(n-r)^2}\right)^{(n-r)}\exp\left(-(n^{(m)}-r^{(m)})E_Q(\theta)\right)\\
&\leq&\exp\left(-(n-r)\left[K(m,n)E_P(\theta)-\log\frac{e}{\epsilon_m}\right]\right)+\exp\left(-(n-r)\left[K(m,n)E_Q(\theta)-\log\frac{(N-n)e^2}{N\epsilon_m^2}\right]\right),
\end{eqnarray*}
here we used the fact that $\epsilon_m\leq \frac{n-r}{n}$ since $r\leq (1-\epsilon_m) n$.

By Lemma \ref{lem0}, (\ref{ep}) and condition (\ref{cond:2}), for a constant $c$, we have
\[C_N:=K(m,n)E_P(\theta)-\log\frac{e}{\epsilon_m}\geq c\eta^2K(m,n)D(P\|Q)-\log\frac{e}{\epsilon_m}\asymp \frac{1}{\epsilon_m^2}.\]
By the fact that $E_Q(\theta)=E_P(\theta)+\theta$ and the definition of $\eta$, one has
\begin{eqnarray*}
D_N&\equiv&K(m,n)E_Q(\theta)-\log\frac{(N-n)e^2}{N\epsilon_m}\\
&\geq& c\eta^2K(m,n)D(P\|Q)-2\log\frac{e}{\epsilon_m}+(1-\eta)K(m,n)D(P\|Q)-\log\frac{N-n}{n}\\
&\geq&c\eta^2K(m,n)D(P\|Q)-2\log\frac{e}{\epsilon_m}\asymp \frac{1}{\epsilon_m^2}.
\end{eqnarray*}

Consequently, 
\[\mathbb{P}[R\leq (1-\epsilon_m) n]\leq \sum_{r=\epsilon_mn}^{\infty}\exp\left(-rC_N\right)+\sum_{r=\epsilon_mn}^{\infty}\exp\left(-rD_N\right)=\exp\left(-\Omega\left(\frac{n}{\epsilon_m}\right)\right).\]

{\bf (Necessity).} For fixed indexes $i,j\in [N]$, let $Z_{ij}=\{Z_k: k\neq i,j\}$. If $Z_i=0$, randomly and uniformly select a node from $\{j:Z_j=1\}$ and denote it as $J$. If $Z_i=1$, randomly and uniformly select a node from $\{j:Z_j=0\}$ and denote it as $J$. Note that
\[\mathbb{P}(J=k|Z_i=0)=\mathbb{P}(J=k|Z_i=0,k\in \{j:Z_j=1\})\mathbb{P}(k\in \{j:Z_j=1\}|Z_i=0)=\frac{1}{N-1}=\mathbb{P}(J=k|Z_i=1),\]
and $\mathbb{P}(Z_{iJ}|Z_i=0,J)=\mathbb{P}(Z_{iJ}|Z_i=1,J)=\binom{N-2}{n-1}^{-1}$.
Then by the property of conditional probability, careful calculation yields
\[
\frac{\mathbb{P}(J,Z_{iJ},A |Z_i=0)}{\mathbb{P}(J,Z_{iJ},A|Z_i=1)}
=\frac{\mathbb{P}(A|Z_{iJ}, J,Z_i=0)}{\mathbb{P}(A|Z_{iJ}, J,Z_i=1)}
\]
\begin{eqnarray*}
&=&\frac{\prod_{\substack{i_2<\dots <i_m\\ i,J\not\in \{i_2,\dots,i_m\}\\
Z_{i_2}\dots Z_{i_m}=1}}Q(A_{ii_2\dots i_m})\prod_{\substack{i_2<\dots <i_m\\ i,J\not\in \{i_2,\dots,i_m\}\\
Z_{i_2}\dots Z_{i_m}=0}}Q(A_{ii_2\dots i_m})\prod_{\substack{i_2<\dots <i_m\\ i,J\not\in \{i_2,\dots,i_m\}\\ Z_{i_2}\dots Z_{i_m}=1}}P(A_{Ji_2\dots i_m})\prod_{\substack{i_2<\dots <i_m\\ i,J\not\in \{i_2,\dots,i_m\}\\ Z_{i_2}\dots Z_{i_m}=0}}Q(A_{Ji_2\dots i_m})}{\prod_{\substack{i_2<\dots <i_m\\ i,J\not\in \{i_2,\dots,i_m\}\\
Z_{i_2}\dots Z_{i_m}=1}}P(A_{ii_2\dots i_m})\prod_{\substack{i_2<\dots <i_m\\ i,J\not\in \{i_2,\dots,i_m\}\\
Z_{i_2}\dots Z_{i_m}=0}}Q(A_{ii_2\dots i_m})\prod_{\substack{i_2<\dots <i_m\\ i,J\not\in \{i_2,\dots,i_m\}\\ Z_{i_2}\dots Z_{i_m}=1}}Q(A_{Ji_2\dots i_m})\prod_{\substack{i_2<\dots <i_m\\ i,J\not\in \{i_2,\dots,i_m\}\\ Z_{i_2}\dots Z_{i_m}=0}}Q(A_{Ji_2\dots i_m})}
\end{eqnarray*}
\begin{eqnarray*}
&&\times \frac{\prod_{\substack{i_1<i_2<\dots <i_m\\ i,J\not\in\{i_1,i_2,\dots,i_m\}\\ Z_{i_1}\dots Z_{i_m}=1}}P(A_{i_1i_2\dots i_m})}{\prod_{\substack{i_1<i_2<\dots <i_m\\ i,J\not\in\{i_1,i_2,\dots,i_m\}\\ Z_{i_1}\dots Z_{i_m}=1}}P(A_{i_1i_2\dots i_m})}\times \frac{\prod_{\substack{i_1<i_2<\dots <i_m\\ i,J\not\in\{i_1,i_2,\dots,i_m\}\\ Z_{i_1}\dots Z_{i_m}=0}}Q(A_{i_1i_2\dots i_m})}{\prod_{\substack{i_1<i_2<\dots <i_m\\ i,J\not\in\{i_1,i_2,\dots,i_m\}\\ Z_{i_1}\dots Z_{i_m}=0}}Q(A_{i_1i_2\dots i_m})}\frac{\prod_{\substack{i_1<\dots <i_m\\ i,J\in\{i_1,\dots,i_m\}}}Q(A_{i_1i_2\dots i_m})}{\prod_{\substack{i_1<\dots <i_m\\ i,J\in\{i_1,\dots,i_m\}}}Q(A_{i_1i_2\dots i_m})}\\
&=&\prod_{\substack{i_2<\dots <i_m\\ i,J\not\in \{i_2,\dots,i_m\}\\
Z_{i_2}\dots Z_{i_m}=1}}\frac{Q(A_{ii_2\dots i_m})}{P(A_{ii_2\dots i_m})}\frac{P(A_{Ji_2\dots i_m})}{Q(A_{Ji_2\dots i_m})}.
\end{eqnarray*}
Let $T_{iJ}=\{A_{ii_2\dots i_m},A_{Ji_2\dots i_m}: i_2<\dots <i_m,i,J\not\in \{i_2,\dots,i_m\},Z_{i_2}\dots Z_{i_m}=1\}$. For $Z_i=0$, $T_{iJ}\sim Q^{\otimes (n-1)^{(m-1)}}\otimes P^{\otimes (n-1)^{(m-1)}}$ and $T_{iJ}\sim P^{\otimes (n-1)^{(m-1)}}\otimes Q^{\otimes (n-1)^{(m-1)}}$ if $Z_i=1$.
Then by a similar argument as in proof of Theorem 1 in \cite{HWX17}, we conclude $ (n-1)^{(m-1)}D(P\|Q)\rightarrow\infty$ and $\liminf_{N\rightarrow\infty}\frac{n^{(m)}D(P\|Q)}{n\log\frac{N}{n}}\geq1$.

\end{proof}

Next, we derive sufficient and necessary conditions for exact recovery of a subhypergraph. Let $E_Q(t)=\sup_{\lambda\in\mathbb{R}}\left(\lambda t-\phi_Q(\lambda)\right)$ be the Fenchel conjugate of $\phi_P$. Obviously, $E_P(t)=\sup_{\lambda\in\mathbb{R}}\left(\lambda t-\phi_P(\lambda)\right)=E_Q(t)-t$. The following theorem presents 
necessary and sufficient conditions for exact recovery of a subhypergraph.

\begin{Theorem}[Exact Recovery]\label{thm:2}
Suppose $P$ and $Q$ satisfy (\ref{cond:1}). If (\ref{cond:2}) and the following hold
\begin{equation}\label{cond:4} 
\liminf_{N\rightarrow\infty}\frac{n^{(m-1)}E_Q\left(\frac{1}{n^{(m-1)}}\log\frac{N}{n}\right)}{\log N}>1,
\end{equation}
then there exists an estimator $\widetilde{S}$ of $S^*$ such that $\mathbb{P}\left(\widetilde{S}=S^*\right)=1+o(1)$.
Conversely, if there exists an estimator $\widetilde{S}$ of $S^*$ such that $\mathbb{P}\left(\widetilde{S}=S^*\right)=1+o(1)$, then (\ref{cond:3}) and the following hold
\begin{equation}\label{cond:5} 
\liminf_{N\rightarrow\infty}\frac{n^{(m-1)}E_Q\left(\frac{1}{n^{(m-1)}}\log\frac{N}{n}\right)}{\log N}\geq1,
\end{equation}
\end{Theorem}

Theorem \ref{thm:2} says that the conditions (\ref{cond:2}) and (\ref{cond:4}) are nearly sharp for exact recovery. 
Again, these conditions are fundamentally different from the ones derived
in the literature of stochastic block models (\cite{ALS18,LKH21}) or generalized censored block models (\cite{ALS19}). 
In particular, exact recovery in \cite{LKH21} only requires a condition similar to (\ref{cond:2}) up to some constant. In contrast, our exact recovery conditions requires additionally (\ref{cond:4}). 
For $m=2$, conditions (\ref{cond:4}) and (\ref{cond:5}) degenerate to Theorem 2 of \cite{HWX17}.

The proof of Theorem \ref{thm:2} heavily relies on Theorem \ref{thm:1}.
Specifically, under (\ref{cond:1}) and (\ref{cond:2}), one can construct a provably exact recovery estimator $\widetilde{S}$ based on weak recovery.
Specifically, one construct $\widetilde{S}$ as follows.
\begin{itemize}
    \item For given $n,N$ and $P,Q, A$, fix a small constant $\delta\in (0,1)$ such that $\delta N,\frac{1}{\delta}$ are integers.
    \item Divide the vertex set $[N]$ into subsets $C_k$ with $|C_k|=\delta N$ for $k=1,2,\dots, \frac{1}{\delta}$.
    \item For $k=1,2,\dots, \frac{1}{\delta}$, let $A_k$ be the subhypergraph on $[N]\setminus C_k $ with $(1-\delta)N$ nodes and subhypergraph of size $(1-\delta)n$. Let $\widehat{S}_k$ be the weak recovery estimator based on $A_k$. 
    \item Let $r_i=\sum_{\substack{i\not\in\{i_2,\dots ,i_m\}\subset \widehat{S}_k,\\ i_2<\dots <i_m}}L_{ii_2\dots i_m}$ for $i\in C_k$ and $k=1,2,\dots, \frac{1}{\delta}$. The exact recovery estimator $\widetilde{S}$ consists of $n$ nodes with the largest $r_i$.
\end{itemize}

\begin{proof}[Proof of Theorem \ref{thm:2}]{\bf (Sufficiency).} 
When $n=O(1)$, exact recover is the same as weak recovery. Hence we assume $n\rightarrow\infty$.
By Lemma \ref{lem4},  we only need to verify (\ref{exacteq1}). Condition (\ref{cond:2}) implies that 
\[\liminf_{N\rightarrow\infty}\frac{((1-\delta)n)^{(m)}D(P\|Q)}{(1-\delta)n\log\frac{(1-\delta)N}{(1-\delta)n}}=(1+o(1))(1-\delta)^{m-1}\liminf_{N\rightarrow\infty}\frac{n^{(m)}D(P\|Q)}{n\log\frac{N}{n}}>1,\]
for small enough fixed constant $\delta\in(0,1)$ and large $n$. Then as argued in the proof of Theorem 2 of \cite{HWX17}, Lemma \ref{lem3} holds. Since $\epsilon_m=o(1)$ in Lemma \ref{lem3} holds. Then  (\ref{exacteq1}) holds.

{\bf (Necessity).} Obviously (\ref{cond:3}) holds. Let $\gamma_m=\frac{1}{n^{(m-1)}}\log\frac{N}{n}$. Then $\gamma_m<\frac{D(P\|Q)}{m(1+\epsilon_0)}$ for a small constant $\epsilon_0>0$. In the following, we only need to prove (\ref{cond:5}) holds. We will prove  (\ref{cond:5}) by contradiction.  If (\ref{cond:5}) is not true, then
\begin{equation}\label{esup}
\limsup_{N\rightarrow\infty}\frac{n^{(m-1)}E_Q\left(\gamma_m\right)}{\log N}<1.
\end{equation}
By Lemma \ref{lem6}, we show there is no $\theta_n$ satisfies (\ref{neceq1}) and (\ref{neceq2}) by showing $\theta_n=\gamma_m$ doesn't satisfy (\ref{neceq1}) or (\ref{neceq2}). There exists a small constant $\tau>0$ such that $\gamma_m$ and $\gamma_m+\tau D(Q\|P)$ are in $[-D(Q\|P), D(P\|Q)]$. Note that for small $\tau$, $(n-1)^{(m-1)}E_Q(\gamma_m+\tau D(Q\|P))<(1-2\tau)\log N$ by (\ref{esup}) and Lemma 2 in \cite{HWX17}. Hence, by Lemma \ref{lem2}, we have
\begin{eqnarray*}
Q\left(\sum_{1\leq i_2<\dots<i_m\leq n-1}L_{i_1i_2\dots i_m}\geq (n-1)^{(m-1)}\gamma_m\right)&\geq& \exp\left(-\frac{(n-1)^{(m-1)}E_Q(\gamma_m+\tau D(Q\|P))+\log2}{1-\frac{C}{(n-1)^{(m-1)}\tau^2\min\{D(Q\|P),D(P\|Q)\}}}\right)\\
&\geq& \frac{1}{n^{1-\tau}}.
\end{eqnarray*}
Then $\gamma_m$ does not satisfy (\ref{neceq2}).

Let $n_1=\frac{n}{\log n}$ and
\[\tau_1:=\frac{(n_1-1)^{(m-1)}D(P\|Q)+[(n-n_1)^{(m-1)}-(n-1)^{(m-1)}]\gamma_m+6\sigma}{(n-n_1)^{(m-1)}D(P\|Q)}=o(1).\]
Take small $\tau>0$ such that $\gamma_m-\tau_1 D(P\|Q)$ and $\gamma_m-(\tau_1+\tau) D(P\|Q)$ are in $[-D(Q\|P), D(P\|Q)]$. 
By (\ref{esup}) and $E_P(\gamma_m)=E_Q(\gamma_m)-\gamma$, for small $\tau$, 
\[(n-n_1)^{(m-1)}E_P(\gamma_m-(\tau_1+\tau) D(Q\|P))\leq (1-2\tau_2)\log n, \]
with a small constant $\tau_2>0$. Then by Lemma \ref{lem2}, it follows
\begin{eqnarray*}
&&P\left(\sum_{1\leq i_2<\dots<i_m\leq n-n_1}L_{i_1i_2\dots i_m}\leq (n-1)^{(m-1)}\gamma_m-(n_1-1)^{(m-1)}D(P\|Q)-6\sigma\right)\\
&=&P\left(\sum_{1\leq i_2<\dots<i_m\leq n-n_1}L_{i_1i_2\dots i_m}\leq (n-n_1)^{(m-1)}(\gamma_m-\tau_1D(P\|Q)\right)\\
&\geq&\exp\left(-\frac{(n-n_1)^{(m-1)}E_P(\gamma_m-(\tau_1+\tau) D(Q\|P))+\log2}{1-\frac{C}{(n-n_1)^{(m-1)}\tau^2\min\{D(Q\|P),D(P\|Q)\}}}\right)\geq \frac{1}{n^{1-\tau_2}}.
\end{eqnarray*}
Hence $\gamma_m$ does not satisfy (\ref{neceq1}).

\end{proof}

\section{Proof of Additional Lemmas}

\subsection{Lemmas for weak recovery}
\begin{Lemma}\label{lem0}
For any fixed integer $m\geq 2$, the following equations hold.
\[K(m,n):=\min_{integer\ r:\ 0\leq r\leq n-1}\frac{n^{(m)}-r^{(m)}}{n-r}=\left(1+o(1)\right)\frac{n^{(m)}}{n},\]
\[\max_{integer\ r:\ 0\leq r\leq n-1}\frac{n^{(m)}-r^{(m)}}{n-r}=\left(1+o(1)\right)(n-1)^{(m-1)}.\]
\end{Lemma}
\begin{proof}
Define a function $f(x)=x(x-1)\dots (x-m+1)$. The derivative of $f(x)$ is equal to 
\[f^{\prime}(x)=\sum_{i=0}^{m-1}x(x-1)\dots (x-i)(x-i-2)\dots (x-m+1).\]
Then $f^{\prime}(x)\geq 0$ for $x\geq m-1$ and $f^{\prime}(x)$ is increasing in $x$. 
Note that $\frac{n^{(m)}-r^{(m)}}{n-r}=\frac{1}{m!}\frac{f(n)-f(r)}{n-r}$. Let $g(x)=\frac{1}{m!}\frac{f(n)-f(x)}{n-x}$. Then
\[g^{\prime}(x)=\frac{1}{m!}\frac{f(n)-f(x)-(n-x)f^{\prime}(x)}{(n-x)^2},\ \ x\leq n-1.\]
By the Lagrange mean value theorem, for $m-1\leq x\leq n-1$, we have
\[\frac{f(n)-f(x)}{n-x}=f^{\prime}(x^*)\geq f^{\prime}(x)\geq0,\ \ x^*\in (x,n).\]
Hence, $g^{\prime}(x)\geq0$ for $x\geq m-1$ and hence $g(r)$ is increasing in $r\geq m-1$. The minimum can only be achieved at $r\in [0, m-1]$. In this case, $\min_{0\leq r\leq m-1}g(r)=\frac{n^{(m)}}{n}(1+o(1))$.
The maximum is achieved at $r=n-1$. By Pascal's rule, we have $n^{(m)}=(n-1)^{(m-1)}+(n-1)^{(m)}$. Then the desired result follows.
\end{proof}

In the following, we present several lemmas from \cite{HWX17}.

\begin{Lemma}[\cite{HWX17}]\label{lem1}
Condition (\ref{cond:1}) implies the following
\begin{eqnarray}\label{ep}
E_P((1-\eta)D(P\|Q))&\geq& \frac{\eta^2}{2C}D(P\|Q),\\ \label{eq}
E_Q(-(1-\eta)D(Q\|P))&\geq& \frac{\eta^2}{2C}D(Q\|P),
\end{eqnarray}
for constant $C$ and $\eta\in [0,1]$ and $D(P\|Q)\asymp D(Q\|P) $.
\end{Lemma}

\begin{Lemma}[\cite{HWX17}]\label{lem2}
Suppose condition (\ref{cond:1}) holds. For $-D(Q\|P)\leq \gamma<\gamma+\delta\leq D(P\|Q)$,
\begin{eqnarray*}
\exp\left(-nE_Q(\gamma)\right)&\geq& Q\left(\sum_{i=1}^nL_i>n\gamma\right)\geq \exp\left(-\frac{nE_Q(\gamma+\delta)+\log2}{1-\frac{C\min\{D(Q\|P),D(P\|Q)\}}{n\delta^2}}\right),
\end{eqnarray*}
where $L_{i}=\frac{dP}{dQ}(X_i)$ with independent $X_i\sim Q$.
\end{Lemma}

For later use, we assume $|S^*|$ is random and unknown and present the following lemmas.

\begin{Lemma}\label{lem03}
If $|n_1-n|\leq \frac{n}{\log n}$, then
\[\min_{integer\ r:\ 0\leq r\leq n_1-1}\frac{n_1^{(m)}-r^{(m)}}{n_1-r}=\left(1+o(1)\right)\frac{n^{(m)}}{n},\ \ \ \ \ \ \max_{0\leq r\leq (1-\epsilon_m)n-\frac{n}{\log n}}\left|1-\frac{n_1-r}{n-r}\right|=o(1),\]
\[\min_{integer\ r:\ 0\leq r\leq n_1-1}\frac{n^{(m)}-r^{(m)}}{n_1^{(m)}-r^{(m)}}\geq \frac{1}{m}\left(1+o(1)\right),\]
 with $\epsilon_m=\left(\min\{\log n, n^{m-1}D(P\|D)\}\right)^{-\frac{1}{2}}$.
\end{Lemma}  

\begin{proof}
By Lemma \ref{lem0} and $|\frac{n_1}{n}-1|\leq \frac{1}{\log n}=o(1)$, we have
\[\min_{integer\ r:\ 0\leq r\leq n_1-1}\frac{n_1^{(m)}-r^{(m)}}{n_1-r}=\left(1+o(1)\right)\frac{n_1^{(m)}}{n_1}=\left(1+o(1)\right)\frac{n}{n_1}\frac{n_1^{(m)}}{n^{(m)}}\frac{n^{(m)}}{n}=\left(1+o(1)\right)\frac{n^{(m)}}{n}.\]
Note that  $0\leq \frac{r}{n}\leq 1-\epsilon_m-\frac{1}{\log n}$.  Then
\begin{eqnarray*}
\max_{0\leq r\leq (1-\epsilon_m)n-\frac{n}{\log n}}\left|1-\frac{n_1-r}{n-r}\right|&\leq&\max_{0\leq r\leq (1-\epsilon_m)n-\frac{n}{\log n}}\frac{n}{n-r}\frac{1}{\log n}=\max_{0\leq r\leq (1-\epsilon_m)n-\frac{n}{\log n}}\frac{1}{1-\frac{r}{n}}\frac{1}{\log n}\\
&\leq & \frac{1}{\epsilon_m\log n+1}=o(1).
\end{eqnarray*}
By Lemma \ref{lem0}, the following holds uniformly for all $r\leq (1-\epsilon_m)n-\frac{n}{\log n}$,
\[\frac{n^{(m)}-r^{(m)}}{n_1^{(m)}-r^{(m)}}=\frac{n^{(m)}-r^{(m)}}{n-r}\left(\frac{n_1^{(m)}-r^{(m)}}{n_1-r}\right)^{-1}(1+o(1))\geq (1+o(1))\frac{n^{(m)}}{n}\left((n_1-1)^{(m-1)}\right)^{-1}=(1+o(1))\frac{1}{m}.\]
\end{proof}

\begin{Lemma}\label{lem3}
Suppose Condition (\ref{cond:1}) holds, $n\rightarrow\infty$,  $\limsup_{N\rightarrow\infty}n/N<1$. If condition (\ref{cond:2}) holds and
\[\mathbb{P}\left(||S^*|-n|\leq \frac{n}{\log n}\right)=1-o(1),\]
then 
\[\mathbb{P}\left(|\widehat{S}_{ML}\Delta S^*|\leq 2n\epsilon_m+3\frac{n}{\log n}\right)=1-o(1), \ \ \ \ \ \epsilon_m=\left(\min\{\log n, n^{m-1}D(P\|D)\}\right)^{-\frac{1}{2}}.\]
\end{Lemma}

\begin{proof}
Let $n_1=|S^*|$. Then $|n_1-n|\leq \frac{n}{\log n}$ with probability tending to 1. 
Let $\widehat{S}=\widehat{S}_{ML}$ and $R=|\widehat{S}\cap S^*|$. Then  $|\widehat{S}\Delta S^*|=n_1+n-2R$. To prove (\ref{weak}), we only need to show $\mathbb{P}\left(R\leq (1-\epsilon_m) n-\frac{n}{\log n}\right)=o(1)$. 

By condition (\ref{cond:2}), there exists a constant $\eta\in (0,1)$ such that $K(m,n)D(P\|Q)\geq (1-\eta)\log\frac{N}{n}$. Let $\theta=(1-\eta)D(P\|Q)$. For $0\leq r\leq n-1$, we have
\begin{eqnarray*}
\{R=r\}&\subset&\{\exists S\subset S^*: |S|=n_1-r, L(S,S^*)\leq (n^{(m)}-r^{(m)})\theta\}\\
&\cup& \{\exists S\subset S^*, \exists T\subset (S^*)^c: |S|=n_1-r,|T|=n-r, L(T,T)+L(T, S^*\setminus S)\geq (n^{(m)}-r^{(m)})\theta\}.
\end{eqnarray*}
Hence, by Lemma \ref{lem03}, we get
\begin{eqnarray*}
\mathbb{P}[R=r]&\leq&
\left(\frac{n_1e}{n_1-r}\right)^{n_1-r}\exp\left(-(n_1^{(m)}-r^{(m)})E_P\left(\frac{n^{(m)}-r^{(m)}}{n_1^{(m)}-r^{(m)}}\theta\right)\right)\\
&&+\left(\frac{n_1e}{n_1-r}\right)^{n_1-r}\left(\frac{(N-n_1)e}{n-r}\right)^{n-r}\exp\left(-(n^{(m)}-r^{(m)})E_Q(\theta)\right)\\
&\leq&\exp\left(-(n-r)(1+o(1))\left[K(m,n)E_P\left(\frac{n^{(m)}-r^{(m)}}{n_1^{(m)}-r^{(m)}}\theta\right)-\log\frac{e}{\epsilon_m}\right]\right)\\
&&+\exp\left(-(n-r)(1+o(1))\left[K(m,n)E_Q(\theta)-\log\frac{(N-n_1)e^2}{n\epsilon_m^2}\right]\right),
\end{eqnarray*}
By Lemma \ref{lem0}, Lemma \ref{lem03}, (\ref{ep}) and condition (\ref{cond:2}), for a constant $C>0$, we have
\[C_N:=K(m,n)E_P\left(\frac{n^{(m)}-r^{(m)}}{n_1^{(m)}-r^{(m)}}\theta\right)-\log\frac{e}{\epsilon_m}\geq (1+o(1))\frac{C}{m}n^{m-1}D(P\|D)-\log\frac{e}{\epsilon_m}\rightarrow\infty.\]
By the fact that $E_Q(\theta)=E_P(\theta)+\theta$, one has
\begin{eqnarray*}
D_N&\equiv&K(m,n)E_Q(\theta)-\log\frac{(N-n_1)e^2}{n\epsilon_m^2}\\
&\geq& c\eta^2K(m,n)D(P\|Q)-2\log\frac{e}{\epsilon_m}+(1-\eta)\frac{n^{(m)}}{n}D(P\|Q)-\log\frac{N-n_1}{n}\\
&\geq&c\eta^2n^{m-1}D(P\|Q)-2\log\frac{e}{\epsilon_m}\rightarrow\infty.
\end{eqnarray*}
Then the proof is complete as in the proof of Theorem \ref{thm:1}.
\end{proof}

\subsection{Lemmas for exact recovery}


\begin{Lemma}\label{lem4}
Let $S_k^*=([N]\setminus C_k)\cap S^*$ for $1\leq k\leq \delta^{-1}$. Suppose condition (\ref{cond:1}) holds and
\begin{equation}\label{exacteq1}
\mathbb{P}\left[|\widehat{S}_k\Delta S_k^*|\leq\delta n, 1\leq k\leq \frac{1}{\delta}\right]=1+o(1).
\end{equation}
Then $\mathbb{P}(\widetilde{S}=S^*)=1+o(1)$.
\end{Lemma}
\begin{proof}
Fix $i_1>(1-\delta n)$ and let $X_{i_1i_2\dots i_m}=\frac{dP}{dQ}(A_{i_1i_2\dots i_m})$ and $Y_{i_1i_2\dots i_m}=\frac{dP}{dQ}(A_{i_1i_2\dots i_m})$ be random variables under $P$ and $Q$ respectively for $i_2<\dots<i_m$.
By a similar argument as in the proof of Theorem 3 in \cite{HWX17}, $r_i$ is stochasticaly greater than or equal to $\sum_{1\leq i_2<\dots<i_m\leq (1-2\delta)n}X_{i_1i_2\dots i_m}+\sum_{1\leq i_2<\dots<i_m\leq \delta n}Y_{i_1i_2\dots i_m}$ for $i\in S^*$. For $i\in [N]\setminus S^*$, $r_i$ is identically distributed as $\sum_{1\leq i_2<\dots<i_m\leq (1-\delta)n}Y_{i_1\dots i_m}$. By Lemma \ref{lem5} below and union bound, $r_i\geq((1-\delta)n)^{(m-1)}\gamma_m, i\in S^*$ and $r_i< ((1-\delta)n)^{(m-1)}\gamma_m, i\in [N]\setminus S^*$ with probability $1+o(1)$. Then $\mathbb{P}(\widetilde{S}=S^*)=1+o(1)$.
\end{proof}

\begin{Lemma}\label{lem5}
Suppose Condition (\ref{cond:1}) holds and let $\gamma_m=\frac{1}{n^{(m-1)}}\log\frac{N}{n}$. Fix $i_1>(1-\delta n)$ and let $X_{i_1\dots i_m}=\frac{dP}{dQ}(A_{i_1\dots i_m})$ and $Y_{i_1\dots i_m}=\frac{dP}{dQ}(A_{i_1\dots i_m})$ be random variables under $P$ and $Q$ respectively for $i_2<\dots<i_m$. Then for small $\delta\in (0,1)$, we have
\[\mathbb{P}\left(\sum_{1\leq i_2<\dots<i_m\leq (1-2\delta)n}X_{i_1\dots i_m}+\sum_{1\leq i_2<\dots<i_m\leq \delta n}Y_{i_1\dots i_m}\leq ((1-\delta)n)^{(m-1)}\gamma_m\right)=o\left(\frac{1}{n}\right),\]
\[\mathbb{P}\left(\sum_{1\leq i_2<\dots<i_m\leq (1-\delta)n}Y_{i_1\dots i_m}\geq ((1-\delta)n)^{(m-1)}\gamma_m\right)=o\left(\frac{1}{N-n}\right).\]
\end{Lemma}

\begin{proof} Condition (\ref{cond:4}) implies $n^{(m-1)}E_Q(\gamma_m)\geq (1+\epsilon)\log N$ for a small constant $\epsilon>0$. Then
\begin{eqnarray*}
&&\mathbb{P}\left(\sum_{1\leq i_2<\dots<i_m\leq (1-\delta)n}Y_{i_1\dots i_m}\geq ((1-\delta)n)^{(m-1)}\gamma_m\right)\leq  \exp\left(-(1-\delta)n)^{(m-1)} E_Q(\gamma_m)\right)\\
&\leq& \exp\left(-(1+o(1))(1-\delta)^{m-1}(1+\epsilon)\log N\right)=N^{-(1-\delta)^{m-1}(1+\epsilon)(1+o(1))}=o\left(\frac{1}{N-n}\right),
\end{eqnarray*}
for $\delta<1-(1+\epsilon)^{-\frac{1}{m-1}}$.

By a similar argument as in the proof of Lemma 5 in \cite{HWX17} and for a constant $C>0$, we have
\begin{eqnarray*}
&&\mathbb{P}\left(\sum_{1\leq i_2<\dots<i_m\leq (1-2\delta)n}X_{i_1\dots i_m}+\sum_{1\leq i_2<\dots<i_m\leq \delta n}Y_{i_1\dots i_m}\leq ((1-\delta)n)^{(m-1)}\gamma_m\right)\\
&\leq& \exp\left(-(1-2\delta)n)^{(m-1)}E_P(\gamma_m)+C(\delta n)^{(m-1)}E_Q(\gamma_m)+(\delta n)^{(m-1)}\gamma_m\right)\\
&\leq & \exp\left\{\left[-n^{(m-1)}E_P(\gamma_m)\left[(1-2\delta)^{m-1}-C\delta^{m-1}\right]+(1+C)\delta^{m-1}\log N\right](1+o(1))\right\}\\
&\leq &\exp\left\{\left[-(\log n+\epsilon \log N)\left[(1-2\delta)^{m-1}-C\delta^{m-1}\right]+(1+C)\delta^{m-1}\log N\right](1+o(1))\right\}\\
&=&\frac{1}{n^{(1-2\delta)^{m-1}-C\delta^{m-1}}}\frac{1}{N^{\epsilon((1-2\delta)^{m-1}-C\delta^{m-1})-(1+C)\delta^{m-1}}}=o\left(\frac{1}{n}\right),
\end{eqnarray*}
if $\delta$ satisfies $\frac{1+(1+C)\delta^{m-1}}{(1-2\delta)^{m-1}-C\delta^{m-1}}-1<\epsilon$. This is possible since 
\[
\frac{1+(1+C)\delta^{m-1}}{(1-2\delta)^{m-1}-C\delta^{m-1}}>1\,\,\,\textrm{and}
\,\,\,
\lim_{\delta\rightarrow 0^+}\frac{1+(1+C)\delta^{m-1}}{(1-2\delta)^{m-1}-C\delta^{m-1}}=1.
\]
\end{proof}

\begin{Lemma}\label{lem6}
Suppose $n\rightarrow\infty$ and $\limsup n/N<1$. Fix $i_1>n$ and let $L_{i_1\dots i_m}=\frac{dP}{dQ}(A_{i_1\dots i_m})$. If there exists an estimator $\widehat{S}$ such that $\mathbb{P}(\widehat{S}=S^*)=1+o(1)$, then for $n_1\rightarrow\infty$ and $n_1=o(n)$, there exists a sequence $\theta_n$ such that 
\begin{eqnarray}\label{neceq1}
P\left(\sum_{1\leq i_2<\dots<i_m\leq n-n_1}L_{i_1i_2\dots i_m}\leq (n-1)^{(m-1)}\theta_n-(n_1-1)^{(m-1)}D(P\|Q)-6\sigma\right)&\leq&\frac{2}{n_1},\\ \label{neceq2}
Q\left(\sum_{1\leq i_2<\dots<i_m\leq n-1}L_{i_1i_2\dots i_m}\geq (n-1)^{(m-1)}\theta_n\right)&\leq&\frac{1}{N-n}.
\end{eqnarray}
Here $\sigma^2=n_1^{(m-1)}Var_P(L_{1\dots m})$ with $Var_P(L_{1\dots m})$ the variance of $L_{1\dots m}$ under $P$.
\end{Lemma}

\begin{proof} Let $\widehat{S}$ be the MLE and $S^*=[n]$. For $i\in S^*$, let $i_0=\argmin_{i\in S^*}L(i,S^*)$. Let $F=\{\min_{i\in S^*}L(i,S^*)\leq \max_{i\not\in S^*}L(i,S^*\setminus\{i_0\}\}$,
\[\theta_{1n}=\inf_{x\in\mathbb{R}}\left\{P\left(\sum_{1\leq i_2<\dots<i_m\leq n-n_1}L_{i_1i_2\dots i_m}\leq (n-1)^{(m-1)}x-(n_1-1)^{(m-1)}D(P\|Q)-6\sigma\right)\geq\frac{2}{n_1}\right\},\]
\[\theta_{2n}=\sup_{x\in\mathbb{R}}\left\{Q\left(\sum_{1\leq i_2<\dots<i_m\leq n-1}L_{i_1i_2\dots i_m}\geq (n-1)^{(m-1)}x\right)\geq\frac{1}{N-n}\right\}.\]
Define $E_1=\{\min_{i\in S^*}L(i,S^*)\leq (n-1)^{(m-1)}\theta_{1n}\}$ and $E_2=\{\max_{i\not\in S^*}L(i,S^*\setminus\{i_0\})\geq (n-1)^{(m-1)}\theta_{2n}\}$.

Suppose $\mathbb{P}(E_1)\geq c_1>0$ and $\mathbb{P}(E_2)\geq c_2>0$ for two constants $c_1,c_2\in (0,1)$. Note that $E_1$ and $E_2$ are independent. Hence, as argued in Lemma 6 of \cite{HWX17}, we have
\[\mathbb{P}(\theta_{1n}>\theta_{2n})\geq \mathbb{P}(E_1\cap E_2\cap F^c)\geq \mathbb{P}(E_1)\mathbb{P}(E_2)-o(1)\geq c_1c_2-o(1)>0.\]
Since $\theta_{1n}$ and $\theta_{2n}$ are deterministic, $\theta_{1n}>\theta_{2n}$ holds for large $n$. Let $\theta_n=\frac{\theta_{1n}+\theta_{2n}}{2}$. Then (\ref{neceq1}) and (\ref{neceq2}) hold.

In the following, we prove $\mathbb{P}(E_1)\geq c_1>0$ and $\mathbb{P}(E_2)\geq c_2>0$ for two constants $c_1,c_2\in (0,1)$. By the right-continuity of $Q$, we have 
\[Q\left(\sum_{1\leq i_2<\dots<i_m\leq n-1}L_{i_1i_2\dots i_m}\geq (n-1)^{(m-1)}\theta_{2n}\right)\geq\frac{1}{N-n}.\]
Then the fact that $1-x\leq e^{-x}$ yields
\begin{eqnarray*}
\mathbb{P}(E_2)&=&1-\prod_{i\not\in S^*}\mathbb{P}\left(L(i,S^*\setminus\{i_0\})< (n-1)^{(m-1)}\theta_{2n}\right)\\
&=&1-\left(1-Q\left(\sum_{1\leq i_2<\dots<i_m\leq n-1}L_{i_1i_2\dots i_m}\geq (n-1)^{(m-1)}\theta_{2n}\right)\right)^{N-n}\geq 1-\frac{1}{e}.
\end{eqnarray*}
Hence $\mathbb{P}(E_2)\geq c_2>0$ for constant $c_2\in (0,1)$. 
By the right-continuity of $P$, we have 
\[P\left(\sum_{1\leq i_2<\dots<i_m\leq n-n_1}L_{i_1i_2\dots i_m}\leq (n-1)^{(m-1)}\theta_{1n}-(n_1-1)^{(m-1)}D(P\|Q)-6\sigma\right)\geq\frac{2}{n_1}.\]
Let $T=[n_1]$ and $T_1=\{i\in T: L(i,T)\leq (n_1-1)^{(m-1)}D(P\|Q)+6\sigma\}$. Then $\mathbb{P}(|T_1|\geq \frac{n_1}{2})=1+o(1)$ by a similar argument as in Lemma 6 of \cite{HWX17}. Hence,
\begin{eqnarray*}
\mathbb{P}(E_1)&\geq& \mathbb{P}\left\{\min_{i\in T_1}L(i,S^*\setminus T)\leq (n-1)^{(m-1)}\theta_{1n}-(n_1-1)^{(m-1)}D(P\|Q)-6\sigma\right\}\\
&\geq&1-\mathbb{P}\left(|T_1|\geq \frac{n_1}{2}\right)\mathbb{P}\left(\prod_{i\in T_1}L(i,S^*\setminus T)\geq (n-1)^{(m-1)}\theta_{1n}-(n_1-1)^{(m-1)}D(P\|Q)-6\sigma \bigg||T_1|\geq \frac{n_1}{2}\right)\\
&&-\mathbb{P}\left(|T_1|< \frac{n_1}{2}\right)\\
&\geq& 1-\exp\left(-\mathbb{P}\left(\sum_{1\leq i_2<\dots<i_m\leq n-n_1}L_{i_1i_2\dots i_m}< (n-1)^{(m-1)}\theta_{1n}-(n_1-1)^{(m-1)}D(P\|Q)-6\sigma\right) \frac{n_1}{2}\right)\\
&&-o(1)\geq 1-\frac{1}{e}-o(1).
\end{eqnarray*}
Then proof is completed.
\end{proof}

\end{document}